\documentclass[conference]{IEEEtran}
\IEEEoverridecommandlockouts
\usepackage{cite}
\usepackage{amsmath,amssymb,amsfonts}
\usepackage{algorithmic}
\usepackage{graphicx}
\usepackage{textcomp}
\usepackage{xcolor}
\usepackage{pgfplots}
\usepackage{xspace}
\RequirePackage{subcaption}

\def\BibTeX{{\rm B\kern-.05em{\sc i\kern-.025em b}\kern-.08em
    T\kern-.1667em\lower.7ex\hbox{E}\kern-.125emX}}

\usepackage{comment}

\newcommand{\sysname}{{\textsc{ROMA}}\xspace}

\begin{document}

\title{\sysname{}: a Read-Only-Memory-based Accelerator for QLoRA-based On-Device LLM
}

\author{Wenqiang Wang$^1$$^2$, Yijia Zhang$^1$$^2$, Zikai Zhang$^2$$^3$, Guanting Huo$^2$$^3$, Hao Liang$^1$$^2$, Shijie Cao$^4$, Ningyi Xu$^*$$^1$$^2$ \\
$^1$Shanghai Jiao Tong University, $^2$Huixi Intelligence, $^3$Peking University, $^4$Microsoft Research Asia\\
$^*$Corresponding email: xuningyi@sjtu.edu.cn
}

\maketitle

\begin{abstract}

As large language models (LLMs) demonstrate powerful capabilities, deploying them on edge devices has become increasingly crucial, offering advantages in privacy and real-time interaction. QLoRA has emerged as the standard approach for on-device LLMs, leveraging quantized models to reduce memory and computational costs while utilizing LoRA for task-specific adaptability.

In this work, we propose \sysname, a QLoRA accelerator with a hybrid storage architecture that uses ROM for quantized base models and SRAM for LoRA weights and KV cache.
Our insight is that the quantized base model is stable and converged, making it well-suited for ROM storage. Meanwhile, LoRA modules offer the flexibility to adapt to new data without requiring updates to the base model.
To further reduce the area cost of ROM, we introduce a novel B-ROM design and integrate it with the compute unit to form a fused cell for efficient use of chip resources.
\sysname can effectively store both a 4-bit 3B and a 2-bit 8B LLaMA model entirely on-chip, achieving a notable generation speed exceeding 20,000 tokens/s without requiring external memory.

\end{abstract}

\begin{IEEEkeywords}
LLM, Accelerator, ROM, SRAM
\end{IEEEkeywords}

\section{Introduction}

Large Language Models (LLMs) have demonstrated remarkable capabilities across a wide range of applications~\cite{zhao2023survey,zhang2022opt,touvron2023llama}.
The potential of deploying LLMs on edge devices, such as autonomous vehicles, robotics, and smartphones, presents exciting opportunities~\cite{li2024large}. 
By processing data locally, LLMs on edge can ensure user privacy, operate in environments with limited or unreliable network connectivity, and deliver faster, real-time responses.
However, deploying LLMs on edge devices presents significant challenges in computation and memory due to the large model size and single-batch autoregressive generation nature of these models.

\begin{figure}
    \centering
    \includegraphics[width=0.9\linewidth]{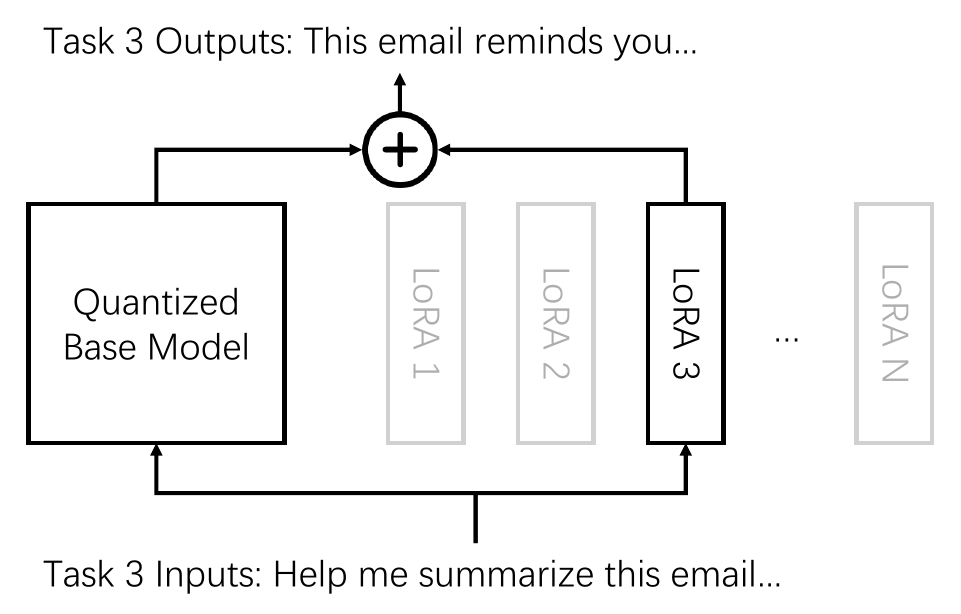}
    \caption{QLoRA, a widely adopted technique for efficient LLM deployment on edge devices.}
    \label{fig:qlora}
\end{figure}

QLoRA has emerged as a leading solution for deploying LLMs on edge devices~\cite{dettmers2024qlora}. Major companies, like Apple and Microsoft, are adopting QLoRA (2/4-bit quantized base model and LoRA adaptors) to power their LLM services on smartphones and desktops~\cite{gunter2024apple,abdin2024phi}. QLoRA combines quantization and Low-Rank Adaptation (LoRA) to optimize model performance while minimizing resource demands~\cite{frantar2022gptq,hu2021lora}. Quantization reduces the memory footprint by lowering the precision of model weights, making it feasible to run large models on devices with limited resources. Meanwhile, LoRA is crucial for adapting foundation models to specific tasks through fine-tuning.
In many edge applications, multiple services or agents require customized LLMs. Instead of deploying a separate large model for each service, QLoRA enables the use of a single, low-bit quantized foundation model, with distinct LoRA modules catering to diverse tasks, as shown in Figure~\ref{fig:qlora}. This approach not only conserves memory but also enhances flexibility and scalability.

Despite its advantages, QLoRA faces several challenges in achieving efficient inference on edge devices. First, although quantization reduces the memory footprint, the QLoRA model still demands significant memory resources. Generating each token requires accessing the entire model weights, where quantized weights dominate in size compared to LoRA weights. Second, QLoRA relies on a mix of low-bit storage and computation for the quantized base model, alongside higher-bit components for LoRA and other operations like attention. Efficiently supporting this mixed-precision computation on hardware introduces a technical challenge, requiring careful design to balance resource allocation between low-bit and high-bit components. 

In this work, we propose \sysname{}, a QLoRA accelerator with a hybrid memory design that leverages ROM for quantized base model and SRAM for LoRA and KV cache.
ROM offers significantly higher memory density compared to SRAM, enabling all quantized base model weights to be stored on-chip. This reduces latency and eliminates the need for external memory access, a critical advantage for efficient inference on edge devices.
A common concern with using ROM is its immutability.
However, we address this with two key insights: (1) Base models are typically stable and converged, as their training approaches consuming almost all available data. The quality improvement becomes incremental over time. We anticipate updates to base models occurring over years, rather than frequent iterations.
(2) While ROM-stored base model weights are fixed, LoRA modules provide the flexibility to incorporate new data and adapt to new tasks without modifying the base model. In summary, combining ROM for low-bit quantized base models with SRAM for LoRA modules provides an effective balance between efficiency and flexibility.

To optimize the area of the ROM, we propose a novel ROM implementation structure named B-ROM. The core idea of B-ROM is to store constant information using the wiring resources on the chip, which reduces the number of CMOS transistors to approximately one-fourth, thus significantly decreasing the area of the ROM. Furthermore, at the physical implementation level, we integrate B-ROM with the compute cell to form a fused cell, which allows for a more efficient use of the chip’s resources and further decreases the overall chip area.

Our hardware implementation demonstrates that \sysname{} achieves a total memory capacity of 1.86 GB ROM, 288 MB SRAM for LoRA and KV cache, and 16 MB SRAM for intermediate data on a single chip, with an area of 503.7mm\(^2\) and power consumption of 33.1W in TSMC's 7 nm technology.
This configuration enables the storage of a 4-bit 3B LLaMA model or a 2-bit 8B LLaMA model entirely in ROM, while supporting a KV cache of up to 4K length and LoRA modules with a rank of 64 in SRAM.
Using the 4-bit 3B LLaMA model with rank-16 LoRA modules as an example, \sysname{} achieves a time-to-first-token (TTFT) latency of 5.6 ms for an input sequence length of 256 and 140.2 ms for a 4K input sequence length. Additionally, the system delivers a decoding throughput of up to 31.8K tokens/s. Even with a 1K KV cached, the decoding throughput remains high at 24.6K tokens/s.
These results highlight \sysname{}'s ability to deliver fast response times and high throughput, demonstrating its feasibility and efficiency for real-time on-device LLM applications.

\section{Background and Motivation}

\subsection{QLoRA: Combining Quantization and Low-Rank Adaptation}
QLoRA is an efficient method for fine-tuning LLMs~\cite{dettmers2024qlora}. It significantly reduces the memory and computation demands of fine-tuning while maintaining model performance by combining low-rank adaptation (LoRA)~\cite{hu2021lora} with low-bit weight quantization~\cite{frantar2022gptq}. 
LoRA introduces additional trainable low-rank matrices into each layer of a pre-trained LLM, updating only these matrices during fine-tuning. Compared to full fine-tuning, this approach drastically reduces the number of trainable parameters. However, LoRA still requires loading the full pre-trained model weights into GPU memory, imposing substantial memory demands.
QLoRA addresses this limitation by integrating LoRA with low-bit weight quantization, storing model weights in lower precision (e.g., INT4) instead of high precision (e.g., FP16), thereby significantly reducing memory usage.
Due to its efficiency and flexibility, QLoRA has become widely adopted for deploying LLMs on edge devices. For example, Apple's Intelligence Foundation Model uses a mix of 2-bit and 4-bit for base model and many LoRA adapters for diverse tasks~\cite{gunter2024apple}.

\subsection{Characteristics and Challenges of QLoRA Inference}

While QLoRA offers advantages for edge device LLM deployment with adaptability to diverse tasks, it faces two key challenges for efficient inference.
First, despite memory savings from quantization, QLoRA still requires significant memory resources. Generating each token necessitates accessing the entire model weights, where the quantized base model dominates in size compared to the lightweight LoRA matrices.
Second, QLoRA relies on mixed-precision computations: low-bit operations for the quantized base model and higher-bit operations (e.g., LoRA and attention). Supporting these mixed precisions efficiently on hardware demands careful resource balancing to avoid performance bottlenecks.

\begin{figure}[htbp]
   \centering
   \includegraphics[width=0.35\textwidth]{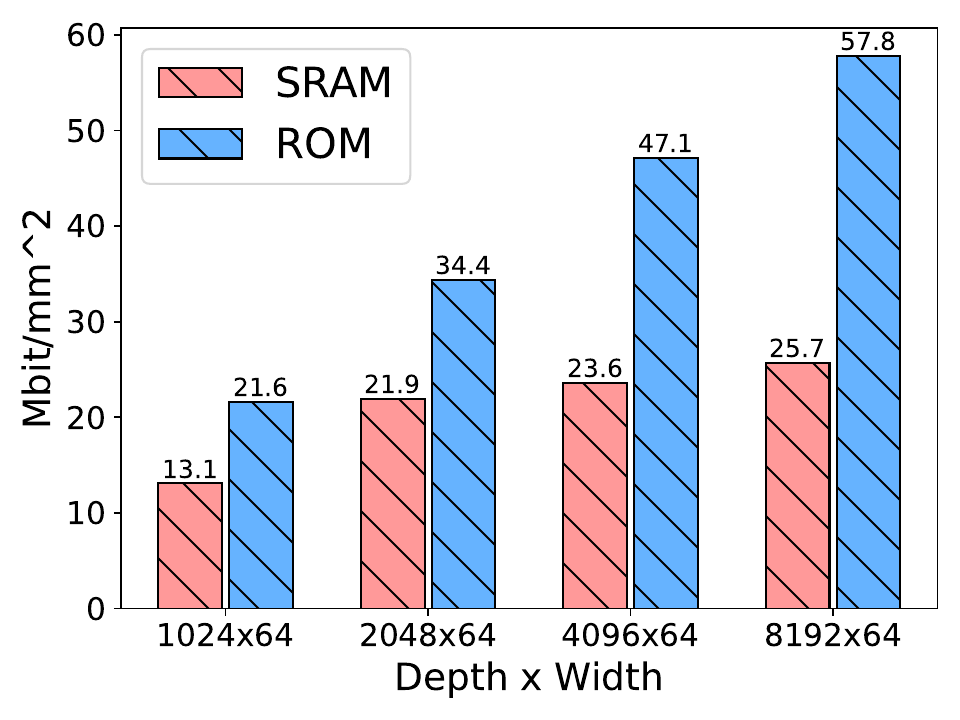}
   \caption{Storage density of SRAM versus ROM.}
   \label{fig:rom_density}
\end{figure}

\begin{figure*}[htbp]
	\centering
	\includegraphics[width=0.7\textwidth]{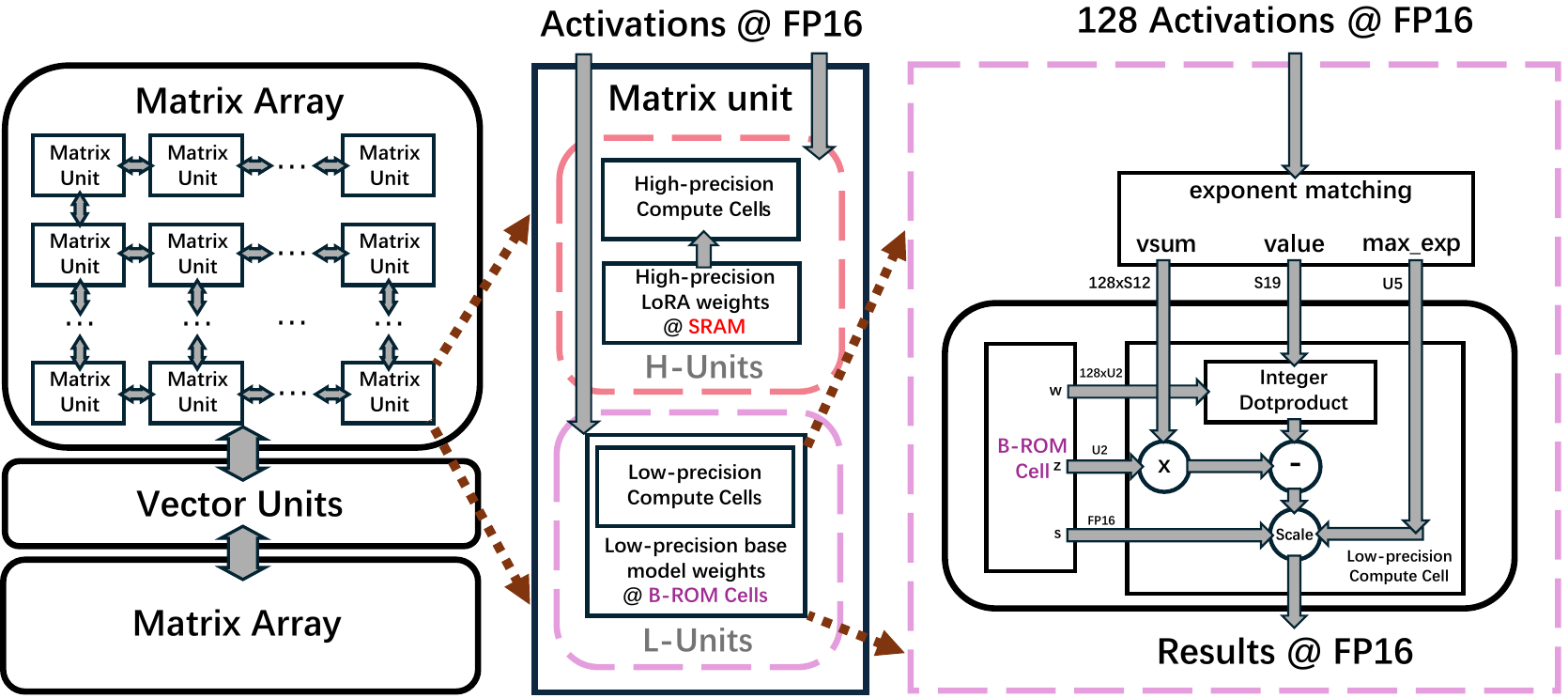}
	\caption{\sysname overall architecture. It is a QLoRA accelerator with hybrid ROM and SRAM.}
    \label{fig_overall}
\end{figure*}

\subsection{Why ROM for Quantized Base Model in QLoRA?}

Read-Only Memory (ROM) is non-volatile memory used to store data that doesn't need modification after manufacturing~\cite{taub1963short,brubaker1975multiplication}. Compared to SRAM, ROM offers significantly higher memory density (as shown in Figure~\ref{fig:rom_density}), making it feasible to store the entire LLM on the chip without external memory.
In conventional designs, each bit of ROM requires only a single CMOS transistor to read information, which occupies much less area compared to the common 6-T transistor structure used in SRAM. Taking the TSMC 7nm Memory Compiler as an example, the area of the bit cell for the generated ROM is approximately one-third of that of the SRAM bit cell, resulting in a higher storage density for the overall ROM cell.

While ROM cannot be modified after it is manufactured, this limitation is not a significant issue for storing the quantized base model for QLoRA on edge devices. Our insight is that the base model is relatively stable and converged, while LoRA provides flexibility to learn new data without updates to the entire base model.
The quantized base model is stable because its training process consumes nearly all available data, with further training only resulting in incremental improvements. As a result, updates to the base model are anticipated to be infrequent, occurring over years rather than rapid iterations.

\section{\sysname Design}

\subsection{Overview: Hybrid ROM and SRAM Architecture}
\label{sec_overall}

As mentioned in Section I, QLoRA is typically used to deploy large language models (LLMs) on edge devices. This means that current on-device deployments of large models typically consist of a foundation model with low-precision general knowledge and multiple high-precision, domain-specific LoRA modules. To store all LLM weights and the KV cache generated during inference on-chip, we propose an LLM accelerator featuring a hybrid ROM and SRAM storage system. 

As illustrated in Figure \ref{fig_overall}, the accelerator is implemented with data flow architecture, consisting of a 2-D computation unit array which is 17 units high and 16 units wide. Each unit can exchange input and results with adjacent units through an internal router. The central part of the accelerator is a 1x16 array of vector units, which is responsible for computing non-matrix operations, including element-wise operations, reduction, and permutation. The vector unit also incorporates a set of SRAM to store intermediate computation results. The two 8x16 arrays located above and below the Vector Unit are composed of Matrix Units, which are designed for matrix-related operations. 

The middle part of Figure \ref{fig_overall} shows the detailed structure of matrix units. Each matrix unit is composed of an SRAM-based H-Unit and a ROM-based L-Unit. The H-Unit is responsible for storing weights for high-precision LoRA modules and KV cache as well as performing related computations. Meanwhile, the L-Units handle the storage and computation of weights for the low-precision foundation model. In the following sections, we focus on the design and optimization of L-Unit because it occupies the largest portion of chip area.

\subsection{L-Unit Design}

The overall structure of the L-Unit is illustrated in the right part of Figure \ref{fig_overall}. The quantized weights are stored in B-ROM (detailedly described in Section III.C), grouped into sets of 128 elements. Each group contains one quantization scale $s$ represented as FP16, one quantization zero-point $z$ represented as UInt2, and a set of quantized values $w$ represented as UInt2. The real value of weight can be calculated as $(w-z)*s$.

The core computational logic of L-Unit involves the matrix-vector multiplication between the quantized weights and input activations. This is done by a shared exponent alignment unit and multiple computation units, as shown in the right part of Figure \ref{fig_overall}. The alignment unit converts the input activations in FP16 format to integer values with a shared exponent, as shown in the following steps: 
\vspace{10pt}

1)\ $vconv = \begin{cases} (-1)^{sign} * \{1'b1,mantissa\}, for\ normal\\ (-1)^{sign} * \{mantissa,1'b0\}, for\ denormal\end{cases}$

2)\ $value_k = vconv_k >> (max\_exp - exp_k)$

\vspace{10pt}
Besides, it  computes the sum of all values, $vsum$, which serves as the input for the subsequent dequantization phase.

After the alignment step, the low-precision compute cells perform the following process to obtain the final dot product result:
\begin{equation}
res = (DotProduct(value,w) - vsum * z) * s * 2^{(max\_exp-Bias)}
\end{equation}

L-Unit optimization is the key to reducing chip area. A straightforward approach involves using a memory compiler to generate standardized ROM modules and combining them with the arithmetic unit, resulting in a total circuit area equal to the sum of the two components. In this paper, we propose two optimizations to further reduce the area. Firstly, we introduce a novel ROM implementation named B-ROM in the L-Unit design, which reduces area by around 40\% compared to standard ROM. Additionally, at the physical design level, we integrate B-ROM cells with compute cells in L-Units, proposing a Fused-cell design. This integration enhances the utilization of on-chip resources, thereby further decreasing chip area. The details will be introduced in the following Sections \ref{sec_brom} and \ref{sec_fuse}.

\subsection{B-ROM: Area-Efficient Block ROM}
\label{sec_brom}

\begin{figure}[htbp]
	\centering
	\includegraphics[width=0.48\textwidth]{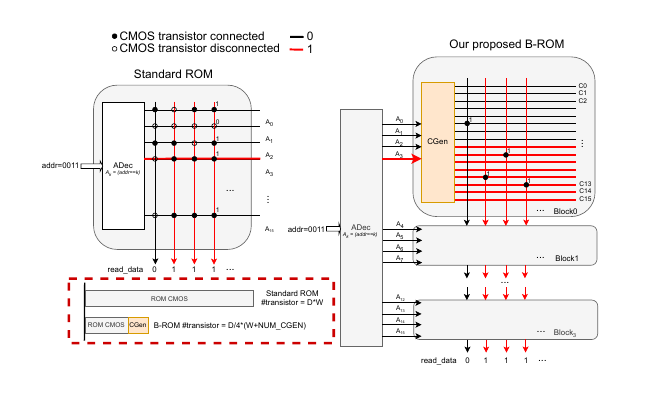}
	\caption{The comparisons between standard ROM and B-ROM. B-ROM need fewer CMOS transistors than ROM.}
    \label{fig_rom}
\end{figure}

A typical ROM structure is composed of an address decoder (ADec) and a storage array~\cite{lewin1965survey}, as shown in the left part of Figure \ref{fig_rom}. Let the depth of the memory cell be $D$ and the width be $W$. The key idea is to translate address to one-hot signals $A_0\sim A_{D-1}$ and then generate outputs with OR operations. Let $M_{i,j}$ represent the content stored in the $j$-th bit of address $i$, then the $j$-th bit output can be expressed by $R_j=\bigvee_{i=0}^{D-1}(A_i\&M_{i,j})$, 
where $\bigvee$ represents the OR operation. At the circuit level, a CMOS transistor is placed at each intersection point of the arrays to realize OR function. If the transistor in the intersection point of $A_i$ and $R_j$ is connected, then it represents the memory content $M_{i,j}=1$. Thus the total number of transistors in the array is $D*W$, which dominates the final area of ROM cell.

Given that the OR operation satisfies the associative property, we can consider grouping the signals $A_0\sim A_{D-1}$, obtaining the output for each group first, and then performing an overall OR operation on these group outputs. Since the connection relationships within each group are finite, we can pre-generate all possible candidates and then directly select from them. We propose a new ROM structure based on this idea, as shown in the right part of Figure \ref{fig_rom}. Firstly, the addresses are grouped into blocks of four. For each block, we pre-generate 16 output candidates $C0$ to $C15$, where $C_k$ represents the output signal when the storage content in the block is $k$. Specifically, for the block with id $b$, $C_k = \bigvee_{i=0}^{3}(A_{4b+i}\&k\_bit\_i)$. At the  circuit level, each block only requires $W$ transistors and the array requires $(\frac{D}{4}*(W + NUM\_CGEN))$ transistors in total, where $NUM\_CGEN$ accounts for the overhead of the CGen module. When $W$ is relatively large, the overhead represented by $NUM\_CGEN$ is significantly less than $W$, which leads to a reduction to approximately 1/4 of the number of CMOS transistors compared to a standard ROM.

\subsection{Fused-Cell: Physical Level Optimization for L-Units}
\label{sec_fuse}

\begin{figure}[htbp]
	\centering
	\includegraphics[width=0.48\textwidth]{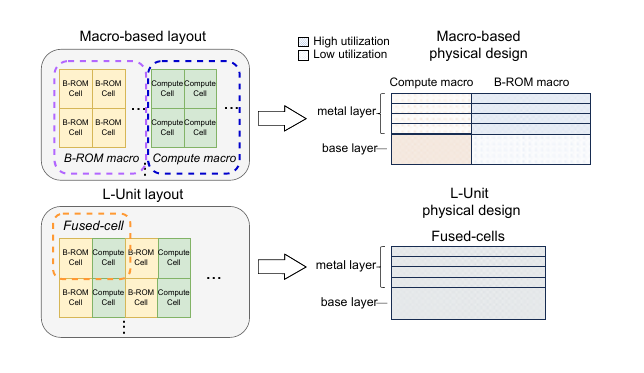}
	\caption{Our Fused-cell layout shows better area efficiency over macro-based one.}
    \label{fig_fusion}
\end{figure}

In conventional CMOS technology, the chip primarily consists of a base layer for CMOS transistors and multiple metal layers for routing and interconnections~\cite{radamson2020state}.
In the design of B-ROM, we significantly reduce the number of CMOS transistors required to store information, but this increases the demand for routing resources. For example, in Figure~\ref{fig_rom}, the number of horizontal interconnections in the B-ROM structure is four times that of a standard ROM. If we generate B-ROM as a macro using memory compilers and analyze its physical design in Figure~\ref{fig_fusion}, we can find that the B-ROM macro's fewer CMOS transistors can lead to low utilization in the base layer, while the larger demand for routing and interconnections leads to high utilization in the metal layers. Conversely, the physical design of our compute macro is dense in the base layer and sparse in the metal layers.

Inspired by the insight that B-ROM macros and compute macros are complementary in physical design, we propose an optimized layout, named Fused-cell. In each Fused-cell, we integrate a B-ROM cell with a compute cell in physical implementation. As shown in the lower part of Figure \ref{fig_fusion}, the two modules are fused within the same physical area, with the B-ROM occupying more of the metal layer and the computation logic occupying more of the base layer. This integration improves the utilization of both base and metal layers, thereby reducing the overall area of the module.

\section{Evaluation}

\begin{table}[htp]
\centering
\begin{tabular}{c|c|c|c} 
\hline
 \textbf{frequency} &  \textbf{memory} &  \textbf{power} & \textbf{area} \\ \hline
500Mhz  &1.86GB(B-ROM)+304MB(SRAM) & 33.1W & 503.7$mm^2$        \\ \hline

\end{tabular}
\caption{Hardware features of ROMA.}
\label{fig_chipfeature}
\end{table}

\subsection{Setup}

In this section, we conduct evaluation experiments on two powerful models, Llama3-8B~\cite{dubey2024llama} and Llama3.2-3B~\cite{llama32}. 
For Llama3-8B, the base model has been quantized into INT2 with FP8 LoRA modules. For Llama3.2-3B, we quantize the base model to  INT4 while keeping LoRA modules with FP8 formats. 
The model sizes of these two QLoRA-based models are both suitable for edge deployments.
During inference, the activations and KV cache are stored with FP16 formats.
We implement and synthesize ROMA in Verilog
to measure latency, area, and power for each
module. We use the Synopsys Design Compiler with the TSMC 7nm standard library for the synthesis, and estimate the power using Synopsys PrimeTime PX. The chip features of ROMA are shown in Table~\ref{fig_chipfeature}.

\subsection{Overall Performance}

\begin{figure}[htbp]
	\centering
	\includegraphics[width=0.48\textwidth]{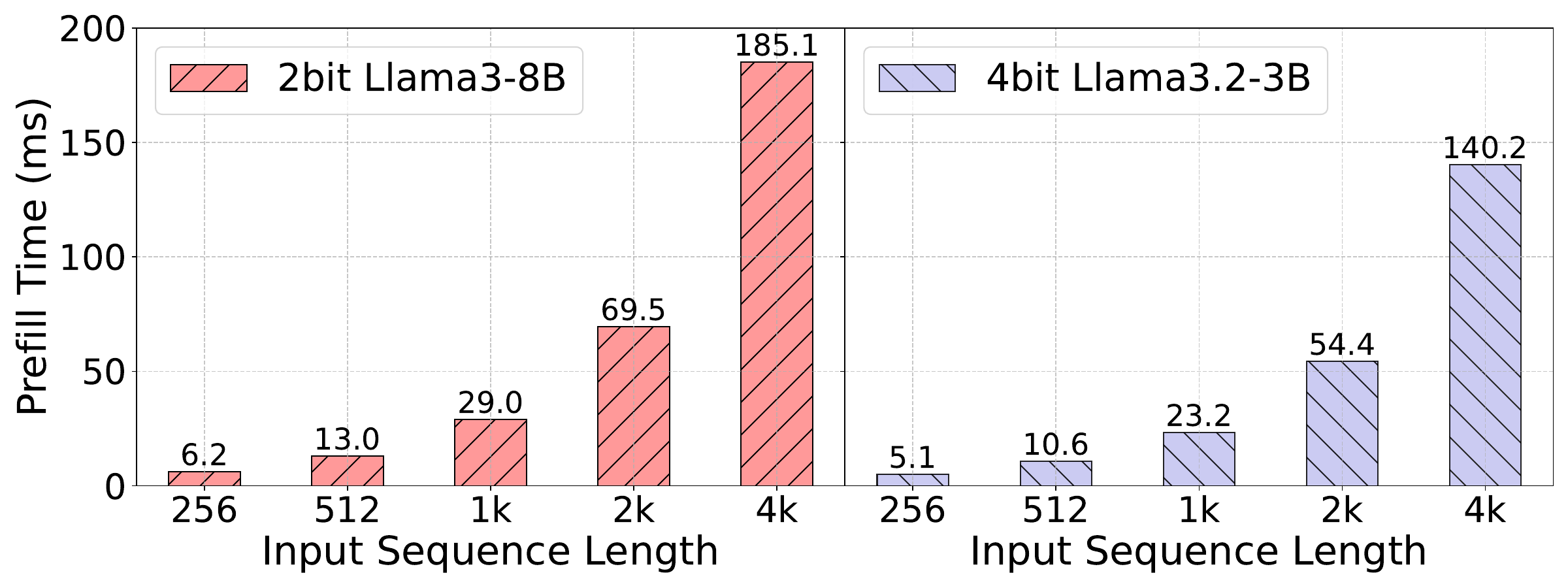}
	\caption{Prefilling time of Llama models on ROMA.}
    \label{fig:prefill}
\end{figure}

\textbf{Prefilling time. }In Figure~\ref{fig:prefill}, we show the prefilling time of QLoRA-based Llama3-8B and Llama3.2-3B models on our ROMA. We keep the QLoRA rank to be 16, and make the input sequence length vary from 256 to 4k. It can be found that the prefilling time can be less than 7ms for both models with input sequence of 256 tokens. As the input sequence length increases, the execution time for prefilling rises, but is still less than 190ms with large sequence length of 4k.

Comparing the overall prefilling performance between the two models,the time required by the 4bit llama3.2-3B model is lower than that of the 2bit llama3-8B model. This is mainly due to the smaller parameter number of the 4-bit-3B model compared to the 2-bit-8B model, leading to faster on-chip memory access. Additionally, the embedding dimension of the 3B model is 3,072, which is smaller than the 4096 of the 8B model, enabling faster computation processing.

\begin{figure}[htbp]
	\centering
	\includegraphics[width=0.48\textwidth]{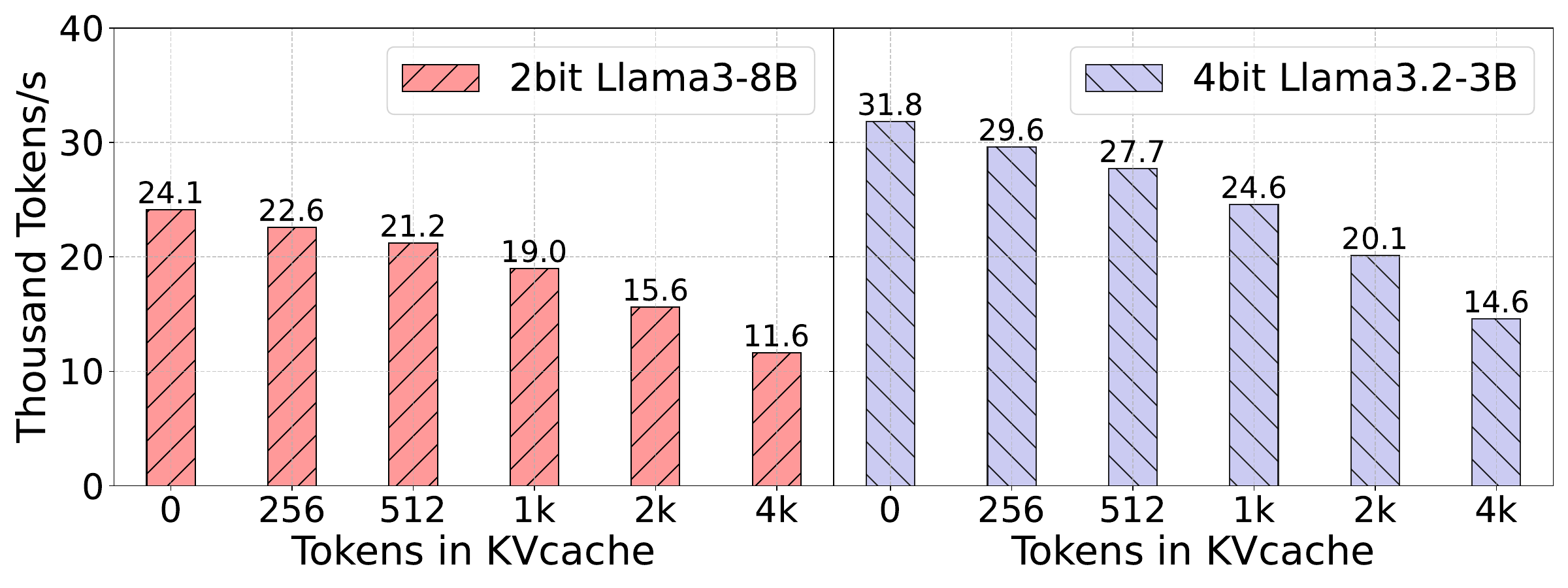}
	\caption{Decoding Performance of Llama models on ROMA.}
    \label{fig:Decode}
\end{figure}

\textbf{Decoding speed. }In Figure~\ref{fig:Decode}, we show the decoding time of QLoRA-based Llama3-8B and Llama3.2-3B models on ROMA. The QLoRA rank is fixed at 16, and the number of tokens accumulated in the KV cache (corresponding to the K and V vectors generated by tokens) varies from 256 to 4k.When the number of accumulated tokens in the KV cache is 0, the peak decoding performance of these two models reaches 24.1k tokens/s and 31.8k tokens/s, respectively. 
Although decoding performance decreases as the number of tokens in the KV cache increases, it remains above 10,000 tokens/s even with 4k tokens.

\begin{figure}[htbp]  
    \centering  
    \begin{subfigure}{0.23\textwidth}  
        \includegraphics[width=\textwidth]{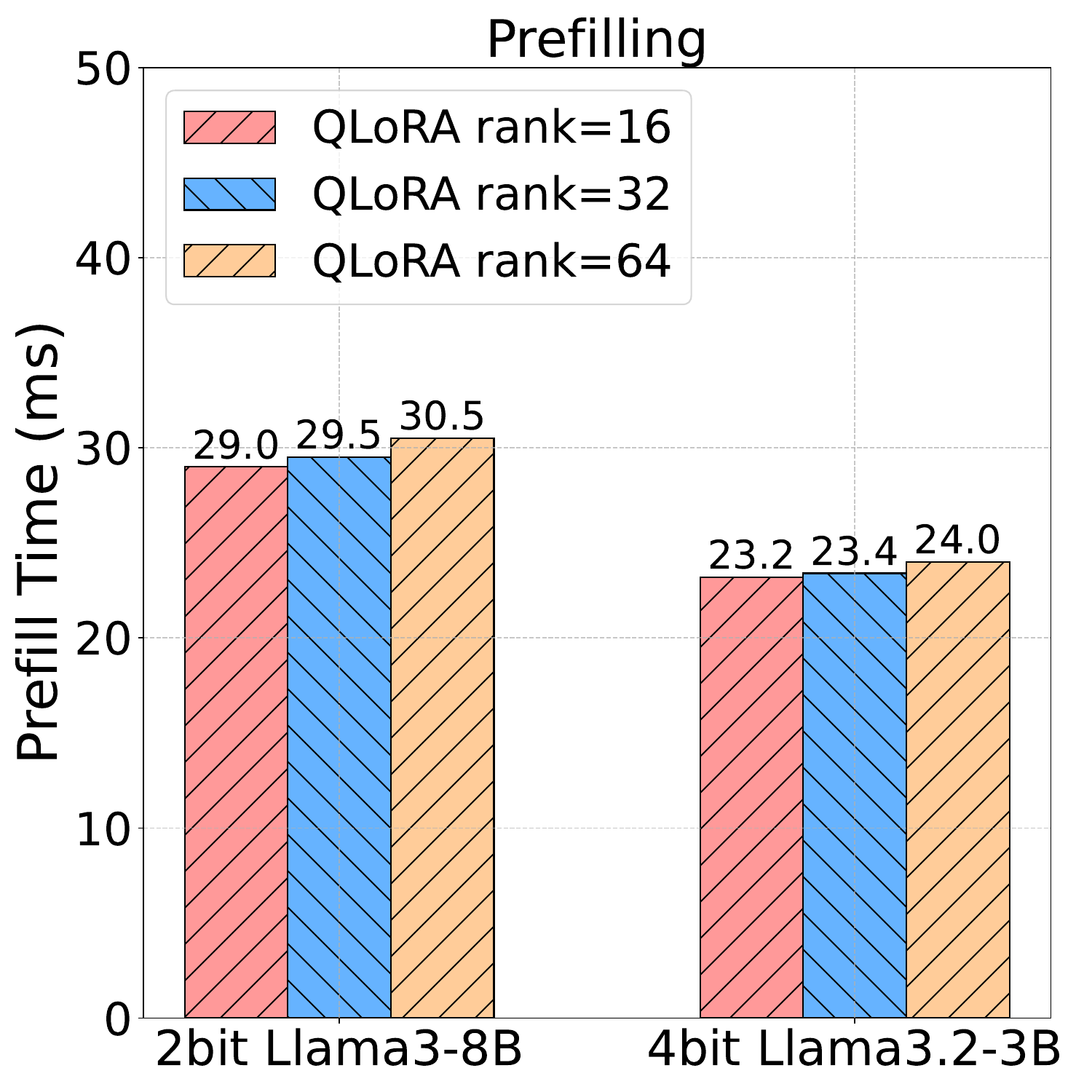}  
    \end{subfigure}  
    \begin{subfigure}{0.23\textwidth}  
        \includegraphics[width=\textwidth]{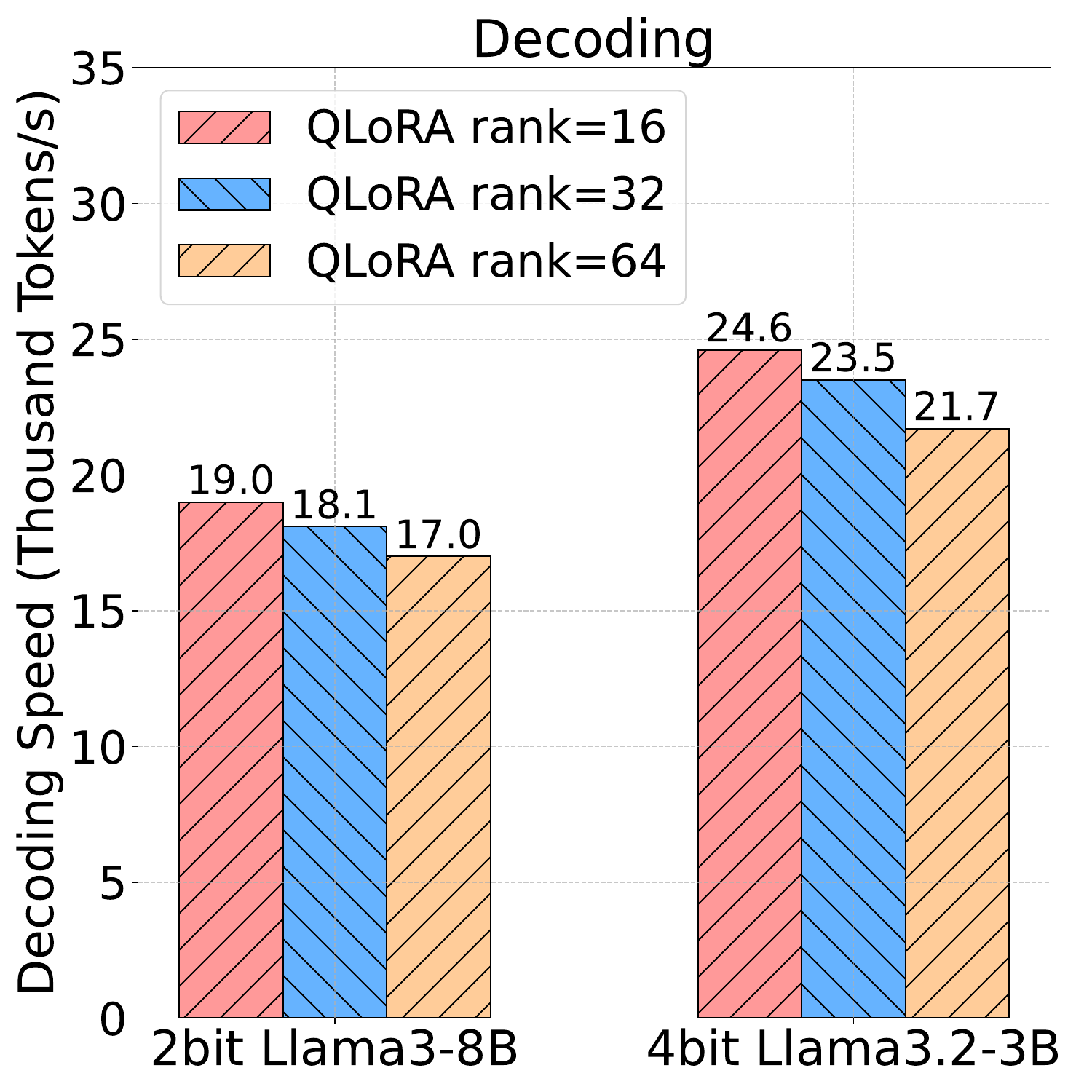}  
    \end{subfigure}  
    \caption{The prefilling and decoding performance of ROMA with different QLoRA ranks.} 
    \label{fig:fpnfbetter} 
\end{figure}

\textbf{The impact of QLoRA ranks. }As the QLoRA rank increases, the prefilling and decoding performance tend to degrade. This phenomenon occurs because the parameter size of QLoRA modules grows with the QLoRA ranks, leading to more matrix-related computation performed by the hardware. In Figure~\ref{fig:fpnfbetter}, we show the impact of QLoRA ranks on the prefilling and decoding performance of two models on ROMA. In the left plot, with the input sequence length fixed at 1k, the prefilling time for both models is under 30ms when the QLoRA rank is set to 16. As the QLoRA rank increases from 16 to 64, the prefilling time increases by less than 5\%. 
The minimal impact of different QLoRA ranks on decoding speed can also be found in the right plot, showing the robustness of ROMA to different ranks.

\textbf{Comparisons with CPU and GPU. }To compare the performance of ROMA with state-of-the-art methods, we evaluate QLoRA-based 4bit Llama3.2-3B on a PC equipped with one Intel i5-1135G7~\cite{i5cpu} and one NVIDIA RTX 4090 GPU~\cite{4090gpu}, denoted as Llama3.2-3B-CPU and Llama3.2-3B-GPU, respectively. These CPU and GPU are common in modern edge devices.
Table~\ref{tab_cpugpu} lists the system configurations for above hardware. 
We use Llama.cpp~\cite{llamacpp} and TensorRT-LLM~\cite{tensorrtllm} frameworks for CPU and GPU LLM inference, respectively. These two frameworks are widely used in edge deployments for LLMs.

\begin{table}[htp]
\centering
\begin{tabular}{c|c|c|c} 
\hline
\textbf{design} & \textbf{frequency} &  \textbf{memory} &  \textbf{power}\\ \hline
CPU: i5-1135G7    & 2.4Ghz  &64GB(DDR)+8MB(SRAM) & 28W        \\ \hline
GPU: RTX 4090     & 2.5Ghz  &24GB(DDR)+73MB(SRAM) & 450W         \\ \hline
\end{tabular}
\caption{Hardware features for the compared chips.}
\label{tab_cpugpu}
\end{table}

\begin{figure}[htbp]
	\centering
	\includegraphics[width=0.48\textwidth]{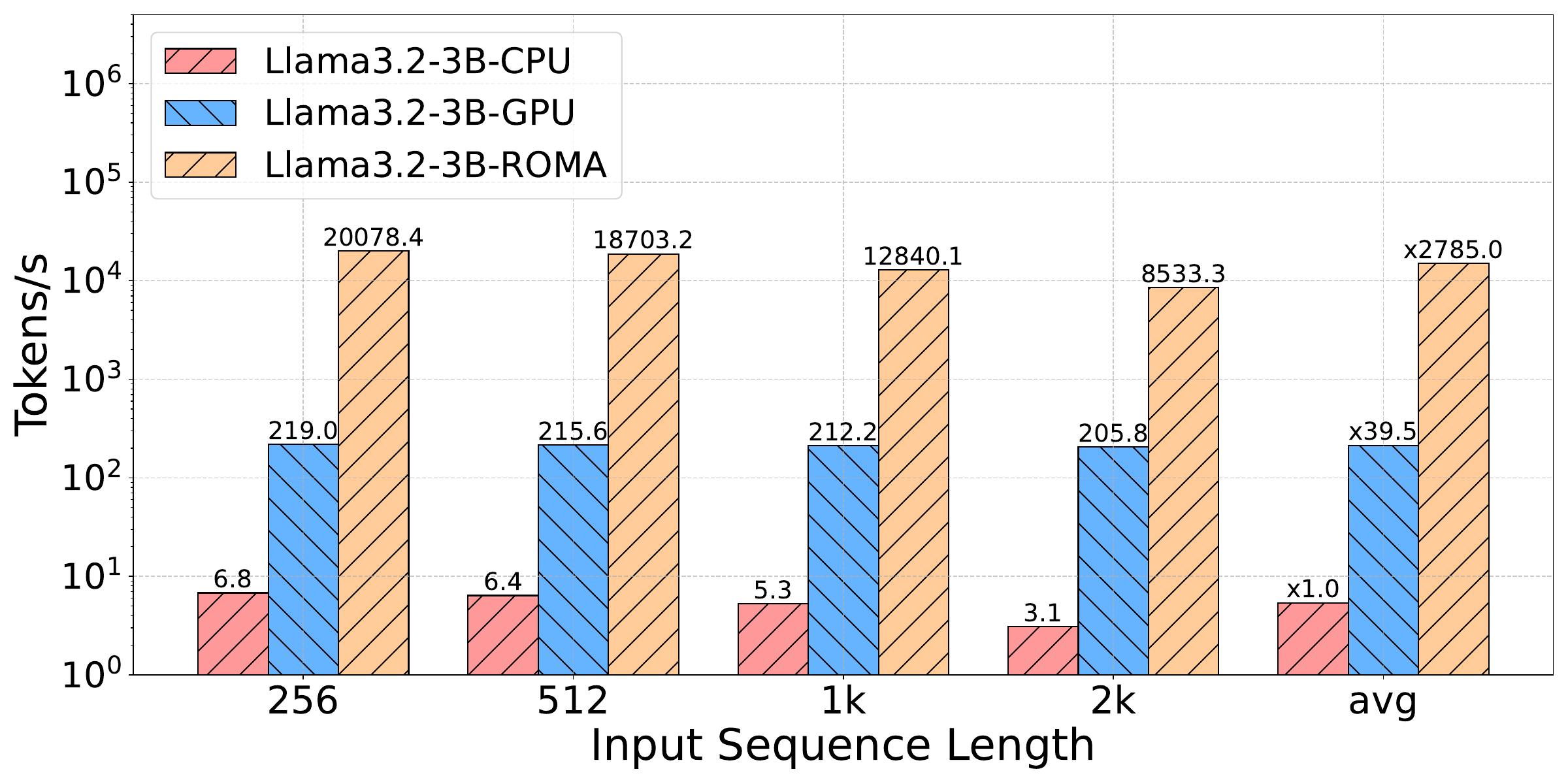}
	\caption{Comparison with edge platforms with CPU and GPU.}
    \label{fig_cpugpu}
\end{figure}

As shown in Figure~\ref{fig_cpugpu}, we fix the output token count at 4k and observe the performance and variations of our ROMA accelerator, RTX 4090, and i5-1135G7 across different input sequence lengths. When the input sequence length is 256 tokens, the CPU achieves 6.8 tokens/s, the GPU achieves 219 tokens/s, while the ROMA accelerator achieves 20,078 tokens/s, demonstrating a significant performance improvement in inference.
As the input sequence length increases from 256 to 2k, the inference performance of all three platforms decreases to some extent. However, even with an input sequence length of 2k tokens, ROMA maintains an impressive performance of 8,533 tokens/s.

By averaging the performance across different input sequence lengths, we find that the inference performance of ROMA is 2,785 times that of the i5-1135G7 CPU, and 70.5 times that of RTX 4090 GPU. This analysis clearly highlights the dramatic performance leap of ROMA in inference tasks over state-of-the-art works.

\begin{figure}[htbp]
	\centering
	\includegraphics[width=0.48\textwidth]{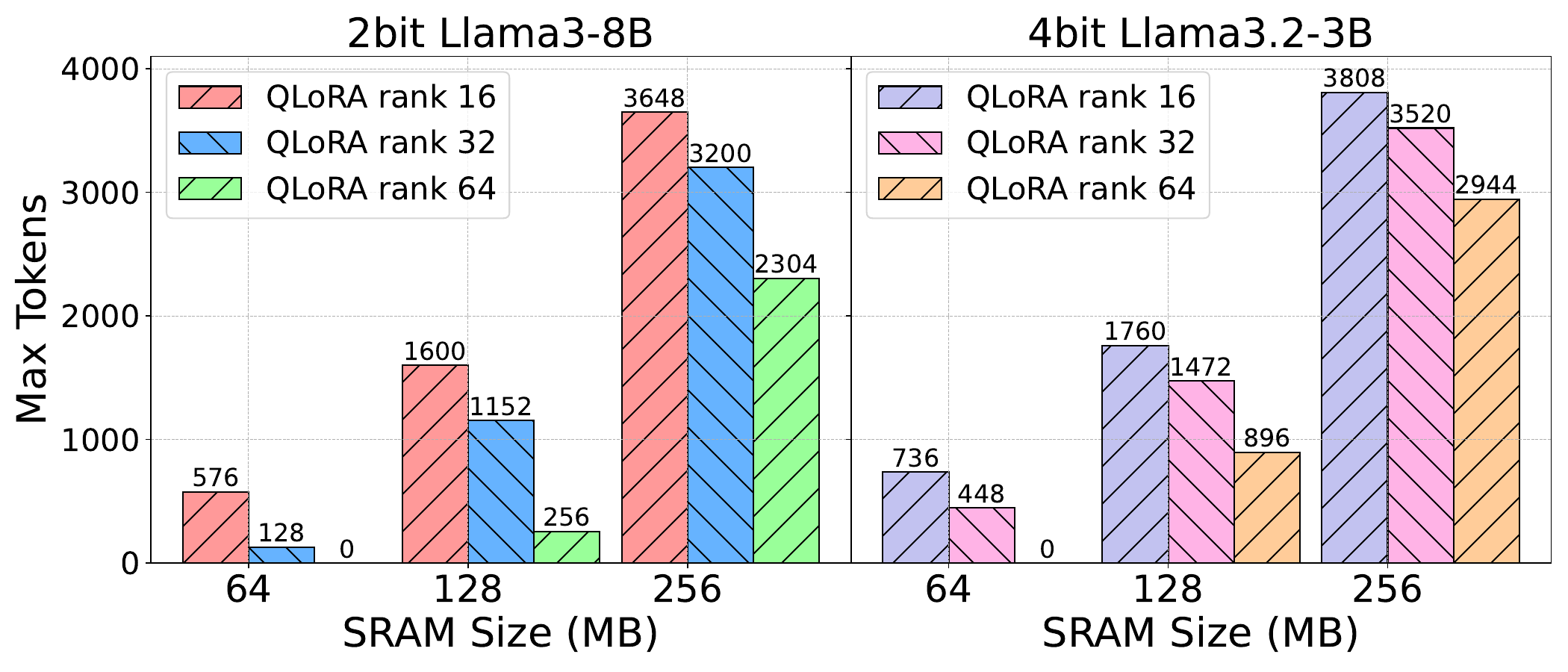}
	\caption{Max tokens ROMA can generate with different SRAM size.}
    \label{fig_maxtoken}
\end{figure}

\textbf{Max token buffer with different SRAM size.}
As shown in Figure~\ref{fig_maxtoken}, we present the maximum number of tokens that can be accommodated by our design under different SRAM capacities. Taking the 4-bit Llama3.2-3B model as an example, when the SRAM capacity is 64 MB and the QLoRA rank is 16, the hardware can accommodate up to 736 tokens. As the SRAM size increases to 256 MB, the maximum token capacity grows to 3,808 tokens, showing good system scalability of ROMA.

We observe that when the SRAM size is 64 MB and the QLoRA rank is 64, the hardware token capacity drops to zero. This occurs because both KV cache data and QLoRA weight data are stored in SRAM. A high QLoRA rank increases the size of the QLoRA weight data, which reduces the space available for the KV cache, thus decreasing the maximum token capacity or even making it impossible to accommodate tokens. This demonstrates that the QLoRA rank negatively impacts the maximum token capacity. Therefore, we set 304MB SRAM in ROMA, making it robust to different rank settings.

\subsection{PPA Analysis}

\begin{figure}[htbp]
	\centering    \includegraphics[width=0.48\textwidth]{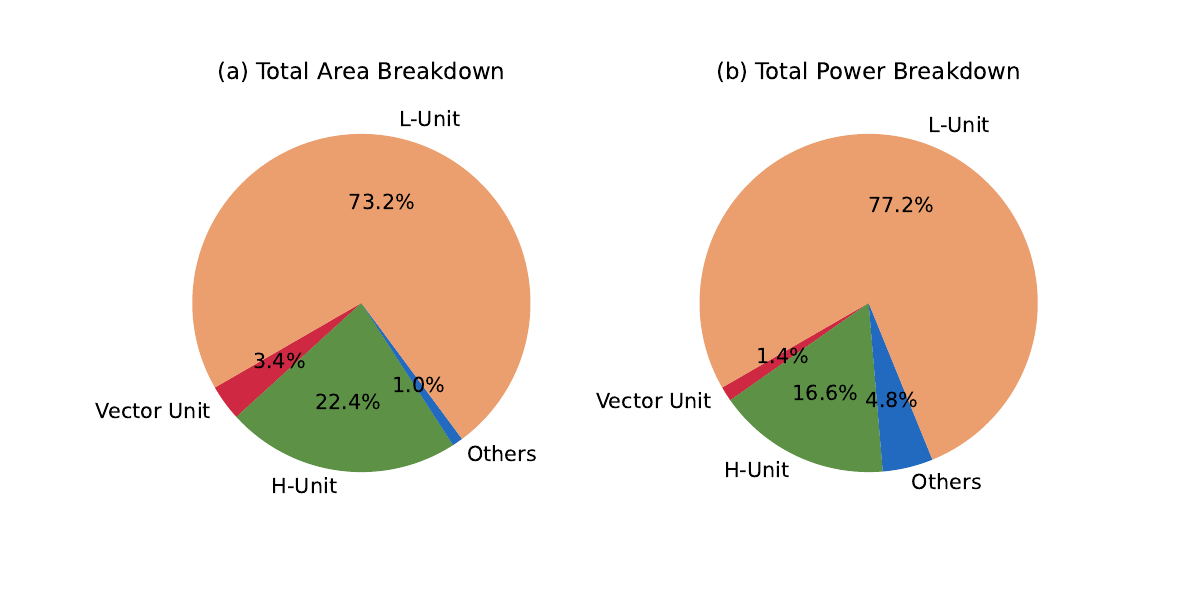}
	\caption{Breakdown of total area and power.}
    \label{fig_ovppa}
\end{figure}

\begin{figure}[htbp]
	\centering    \includegraphics[width=0.34\textwidth]{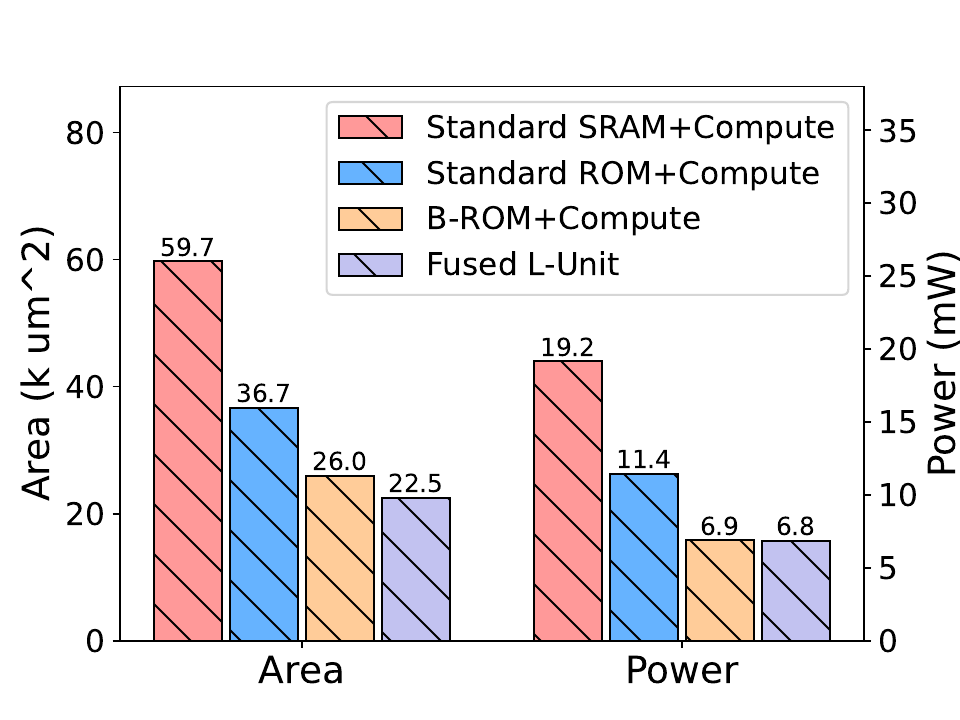}
	\caption{Area and power of L-Unit.}
    \label{fig_muppa}
\end{figure}

Figure \ref{fig_ovppa} shows the breakdown of our overall area and power consumption, in which the L-Unit and H-Unit account for the largest proportions in the entire system. Correspondingly, we have specifically analyzed the benefits brought by various optimizations of the L-Unit. Figure \ref{fig_muppa} compares the area and power consumption of the L-Unit under different designs: Standard SRAM + Compute Cell, Standard ROM + Compute Cell, B-ROM + Compute Cell, and Fused Cell.The adoption of standard ROM results in a significant reduction in both area and power consumption compared to SRAM. Furthermore, with the introduction of our proposed B-ROM and Fused Cell, the overall area and power consumption are further optimized.

\section{Conclusion}

In this work, we present \sysname, a QLoRA accelerator designed to efficiently deploy large language models on edge devices. By leveraging a hybrid memory architecture that combines ROM for quantized base models and SRAM for LoRA weights and KV cache, \sysname addresses the challenges of memory and computational efficiency inherent in on-device LLM inference.
Our hardware implementation demonstrates that \sysname can store QLoRA models without external memory, such as the 4-bit 3B or 2-bit 8B LLaMA models, while achieving impressive prefilling and decoding speeds exceeding 20,000 tokens/s.
These results highlight the potential of \sysname as a powerful and efficient solution for real-time, on-device LLM applications.

\bibliographystyle{IEEEtran}
\bibliography{refs}

\end{document}